\documentclass[fleqn,usenatbib]{mnras}

\DeclareSymbolFont{cmletters}{OML}{cmm}{m}{it}
\DeclareMathSymbol{v}{\mathalpha}{cmletters}{"76}

\usepackage{ae,aecompl}
\usepackage{times}
\usepackage{soul}

\usepackage{tabularx,ragged2e,booktabs,caption}
\usepackage{amsmath}
\usepackage{amssymb}
\usepackage{epsfig}
\usepackage{graphicx}
\usepackage{ifthen}
\usepackage{latexsym}
\usepackage{rotating}
\usepackage{times,epsf}
\usepackage{txfonts}
\usepackage{varioref}
\usepackage{verbatim}
\usepackage{array}
\usepackage{url}
\usepackage{color}
\usepackage[dvipsnames]{xcolor}
\usepackage[T1]{fontenc}
\usepackage{epstopdf}
\usepackage{ulem}

\newcommand{\be}{\begin{equation}}
\newcommand{\ee}{\end{equation}}
\newcommand{\bea}{\begin{eqnarray}}
\newcommand{\eea}{\end{eqnarray}}

\newcommand{\koral}{\texttt{KORAL}}
\newcommand{\MEdd}{\dot M_{\rm Edd}}
\newcommand{\Medd}{\dot M_{\rm Edd}}

\newcommand{\Rg}{R_{\rm g}}

\title[Cosmic battery in accretion flows]{Numerical simulations of the Cosmic Battery in accretion flows around astrophysical black holes}
\author[I. Contopoulos, A. Nathanail, D., A. S\k{a}dowski,  Kazanas, R. Narayan]
       {I. Contopoulos$^{1,2}$\thanks{E-mail: icontop@academyofathens.gr (IC)}, A. Nathanail$^{3,4}$, A. S\k{a}dowski$^{5,6,7}$, D. Kazanas$^{8}$, R. Narayan$^{9}$ \\ 
$^1$ Research Center for Astronomy and Applied Mathematics, Academy of Athens, Athens 11527, Greece\\
$^2$ National Research Nuclear University (MEPhI), Moscow 115409, Russia\\    
$^3$ Institute for Theoretical Physics, D-60438 Frankfurt, Germany\\
$^4$ Humboldt Fellow\\
$^5$ MIT Kavli Institute for Astrophysics and Space Research,
77 Massachusetts Ave, Cambridge, MA 02139, USA\\
$^6$ Einstein Fellow\\
$^7$ Present address: Akuna Capital, 585 Massachusetts Ave., MA 02139, USA\\
$^8$ Astrophysics Science Division, NASA/Goddard Space Flight Center, Greenbelt, MD 20771, USA\\
$^9$ Harvard-Smithsonian Center for Astrophysics, Cambridge, MA 02138, USA
}

\begin{document}

\maketitle

\label{firstpage}

\begin{abstract}
We implement the \koral\ code to perform two sets of very long general relativistic radiation magnetohydrodynamic simulations of an axisymmetric optically thin magnetized flow around a non-rotating black hole: one with a new term in the electromagnetic field tensor due to the radiation pressure felt by the plasma electrons on the comoving frame of the electron-proton plasma, and one without. The source of the radiation is the accretion flow itself. Without the new term, the system evolves to a standard accretion flow due to the development of the magneto-rotational instability (MRI). With the new term, however, the system eventually evolves to a magnetically arrested state (MAD) in which a large scale jet-like  magnetic field threads the black hole horizon. Our results confirm the secular action of the Cosmic Battery in accretion flows around astrophysical black holes.

\end{abstract}


\section{Introduction}
\label{s.introduction}

Magnetic fields are an important constituent of cosmic plasmas at all astrophysical scales. Absent in a homogenous and isotropic universe, most researchers believe that they originate in `battery' currents \citep{B50} that generate seed magnetic fields which are subsequently amplified by stretching in planetary and stellar interiors, accretion disks and generally rotating astrophysical plasmas. Most direct observations through Faraday rotation, polarization, Zeeman effect, etc. show a generally turbulent magnetic field structure on all astrophysical scales, from planets to galaxy clusters.

On the other hand, an important class of astrophysical structures, namely winds and jets from accretion disks around compact objects (primarily black holes), can most economically be explained through large scale grand design helical magnetic fields that extend from the immediate vicinity of the central compact object up to several million times larger scales. This theoretical picture is supported both by indirect observational evidence (e.g. AGN warm absorbers, Faraday Rotation-Measure gradients in astrophysical jets, polarization measurements, XRB Hardness-Intensity Q-diagrams, momentum vs energy content in the wind, GRB exponential afterglow decay, etc.), and by state of the art magneto-hydrodynamic (MHD) simulations. However, there is still no universally accepted theory that may account for the origin of a large scale ordered field near the center of an accretion disk, and several researchers have recently begun to look to other possible mechanisms for astrophysical jets and winds like radiation pressure, hydrodynamics, collimation by an external medium, etc. \citep[e.g.][]{SN15}.

A few years back, a simple astrophysical mechanism was proposed that operates in the innermost accretion flows around astrophysical black holes, the so called `Cosmic Battery' \citep[hereafter CB;][]{CK98}. According to the CB, the electrons of the orbiting plasma at the inner edge of the accretion disk around the central black hole are subject to the radiation pressure of the ambient radiation field, while this is not the case with the plasma protons, on which the radiation pressure is negligible. 
In the frame of the plasma, the radiation field is abberated due to the orbiting motion and therefore, the resulting radiation pressure develops a significant toroidal component in the direction opposite to the direction of rotation\footnote{As we will discuss below, this statement is modified when the central black hole rotates at more than about $70\%$ of maximal rotation. In that case, the inner edge of the accretion disk approaches close enough to the black hole horizon for the space-time rotation effect on photons to win over the abberation due to the orbital motion.}. This is analogous to the well known Poynting-Robertson radiation drag on dust grains in orbit around the sun in our solar system. 
This toroidal component of the radiation force, then, gives rise to a non-zero component of a toroidal electric field $E^\phi$, the curl of which gives rise to a secular growth of poloidal magnetic field.


It is straightforward to obtain an estimate of the timescale for growth, in particular, how much time is required for the CB-induced magnetic field to grow to the same level as the MRI-sustained magnetic field. We make at this stage the simplifying assumption of negligible accretion, i.e., we presume that the CB-related growth takes place in an orbiting non-accreting plasma flow. In hot accretion flows the magnetic pressure saturates at some fraction of the thermal pressure. This, in turn, is close to the virial energy obtained by the gas in a hot accretion flow \citep{YN14}, i.e., $p_{\rm thermal}\approx (\Rg/R) \rho c^2$. The density $\rho$ can be related to the accretion rate, $\dot M\approx 4\pi R^2\rho v_{\rm R}$, assuming the infall velocity to be close to its free-fall value $v_{\rm R}\approx c\sqrt{2\Rg/R}$. Hereafter, we define the gravitational radius $R_{\rm g}\equiv GM/c^2$  for a~black hole of mass $M$ as the unit of length, and $\tau_{\rm g}\equiv R_{\rm g}/c$ as the unit of time. Under such assumptions we obtain an estimate of the magnetic field strength in an optically thin disk as follows,
\be
\label{e.Bestimate}
B_{\rm sat}\approx 
10^8 \left(\frac{R}{\Rg}\right)^{-\frac{5}{4}} \left(\frac{\dot M}{\MEdd}\right)^{\frac{1}{2}} \left(\frac{M}{10M_\odot}\right)^{-\frac{1}{2}} \rm \, G.
\ee
In this work we adopt the following definition for the Eddington mass accretion rate,
\be
\label{e.medd}
\Medd = \frac{L_{\rm Edd}}{c^2},
\ee
where $L_{\rm Edd}=4\pi GMm_{\rm p} c/\sigma_{\rm T}=1.26 \times 10^{38}  M/M_{\odot}\,\rm erg/s$ is the Eddington luminosity. $\sigma_{\rm T}$ is the Thompson cross-section for scattering on electrons, $m_{\rm p}$ is the proton mass, and $c$ is the speed of light. For the case considered in this paper, the spin of the black hole is zero, $M=10M_{\odot}$, and $\Medd = 1.40 \times 10^{18}\,\rm g/s$. 

In an isotropic radiation field of central luminosity $L$ the Keplerian rotation will result in a toroidal component of the radiation energy flux in the flow (comoving) frame
\be 
\label{isotropicrad}
F^{\phi}=\frac{L}{4\pi R^2}\frac{v_{\rm \phi}}{c}\approx \frac{L}{4\pi R^2}\left(\frac{R}{\Rg}\right)^{-\frac{1}{2}}\ .
\ee
The induced electric field equals $E^{\phi}=\sigma_{\rm T} F^{\phi} / c e$, where $e$ is the magnitude of the electron charge. This yields the CB-related growth of the magnetic field, 
\be
\label{e.dBdt.estimate}
\frac{{\rm d}B^{\theta}}{{\rm d}t}= \frac{c}{R}\frac{\rm d}{{\rm d}R}\left(R E^{\phi}\right)
\approx 
10^4 \left(\frac{L}{L_{\rm Edd}}\right) \left(\frac{R}{\Rg}\right)^{-\frac{7}{2}} \left(\frac{M}{10 M_\odot}\right)^{-2} \rm \, G\cdot s^{-1}.
\ee
The CB growth timescale is therefore,
\be
\label{e.battau.estimate}
\tau_{\rm CB}\sim
\frac{B_{\rm sat}}{{\rm d}B^{\theta}/{\rm d}t} \approx 4\times 10^{3}
\left(\frac{L}{L_{\rm Edd}}\right)^{-1}  
\left(\frac{M}{10 M_\odot}\right)^{\frac{3}{2}}
\left(\frac{\dot M}{\MEdd}\right)^{\frac{1}{2}} 
\left(\frac{R}{\Rg}\right)^{\frac{9}{4}}  \rm \, s\ .
\ee
Obviously, in strongly sub-Eddington accretion flows, the timescales required for magnetic field saturation are much longer.

The above introduction on the Cosmic Battery is overly simplified, and its actual implementation is more complicated than our simple growth rate calculation above. In fact, the induced toroidal electric field is confined to a region of small radial extent around the inner edge of the accretion disk, and as a result, it generates poloidal magnetic field loops with one polarity crossing the inermost accretion flow, and the return polarity crossing the disk further out. Obviously, the total generated poloidal magnetic flux that threads the disk is zero. An ideal accretion flow would thus advect these poloidal magnetic field loops toward the vicinity of the central black hole, and as a result, the total accumulated magnetic flux would saturate to a value on the order of $\tau_{\rm g}({\rm d}B^{\theta}/{\rm d}t)$ which is many orders of magnitude smaller than $B_{\rm sat}$ ($\tau_{\rm g}\equiv R_g/v_R\ll \tau_{\rm CB}$). This has been the central point of criticism by \citet{BLB02}. 
What the latter authors did not take into account was that accretion disks are not ideal throughout, except for their inermost section where matter freely falls onto the black hole horizon. They are turbulent, thus also viscous and resistive/magnetically diffusive. Therefore, it is natural to assume that the inner part of the poloidal magnetic field loops is advected by the inermost ideal accretion flow, whereas the outer part of the loops diffuses outward through the turbulent disk. Thus, the two polarities of the generated poloidal magnetic field are separated, the battery mechanism operates continuously, and the field can grow to its maximum possible saturation value. This important point of \citet{CK98} has been missed by \citet{BLB02}.

In order to convince ourselves and the scientific community that the CB operates as described above, we performed in the past several numerical simulations of the process \citep{CKC06, CCK08, CNK15}. In all of them, though, the source of radiation has been considered to be isotropic as in  eq.~(\ref{isotropicrad}) (e.g. a non-rotating central star). Moreover, previous simulations were performed in Newtonian (not general-relativistic) spacetime. 
\citet{KC14} were the first to study the radiation field in an actual accretion flow around a real astrophysical black hole when the source of radiation is the accretion disk itself, and its radiation field is distorted both by the disk and spacetime rotation. We will come back to the main conclusions of their work in \S~4 below. 

In the present work we perform the first fully general-relativistic
magneto-hydrodynamic numerical simulations with radiation that include
the effect of the CB. We consider a magnetized flow around a
$10M_\odot$ non-rotating (Schwarzschild) black hole in which the
radiation field is generated by the flow itself. The dynamical time
$\tau_{\rm g}$ is on the order of a tenth of a millisecond
($10^{-4}$~s). The radiation field in our simulation is strongly
sub-Eddington with $L/L_{\rm Edd}\sim 10^{-4}$, and
eq.~\ref{e.battau.estimate} yields a rough estimate of the timescale
required for the Cosmic Battery to manifest itself, namely, \be
\tau_{\rm CB}\sim 10^8\ \rm{s}\sim 10^{12}\ \tau_{\rm g}\ .  \ee
Obviously, it is impractical to perform a simulation that
lasts for trillions of dynamical times, and therefore, in order to be
able to observe the growth of the magnetic field to saturation within
our finite available simulation time, we have artificially increased
the CB radiation-induced term in the induction equation by nine orders
of magnitude, similarly to what we did in past numerical simulations
of the process. Our results confirm the operation of the CB in
realistic accretion flows.

\section{Numerical simulations}
\label{s.efficiency}

We performed numerical simulations with the general relativistic radiation MHD code \koral\ which employs the M1 scheme to close the radiation moment equations. Essentially, the code treats radiation as a `fluid of photons', and the M1 scheme assumes that there exists a `rest frame' in which the radiation flux vanishes and the radiation tensor is isotropic. For a detailed description of the code, the reader should consult \cite{sadowski+koral,sadowski+koral2}. In the present section, we focus only on the modifications needed to account for the effect of the Cosmic Battery.

The conservation laws for gas density, energy and momentum,
radiation energy and momentum, and photon number can be written in covariant form,
\begin{align}
(\rho u^\mu)_{;\mu}&= 0 ,\label{eq.cons1} \\
(T^\mu_{\ \nu})_{;\mu}&= G_\nu,\label{eq.cons2}\\
(R^\mu_{\ \nu})_{;\mu}&= -G_\nu,\label{eq.cons3}\\
(n u^\mu)_{;\mu}&= \dot n,  \label{eq.cons4}
\end{align}
where $\rho$ is the gas
density in the comoving fluid frame, $u^\mu$ is the gas four-velocity,
$T^\mu_\nu$ is the
MHD stress-energy tensor,
\be\label{eq.tmunu}
T^\mu_{\ \nu} = (\rho+u_{\rm int}+p+b^2)u^\mu u_\nu + (p+\frac12b^2)\delta^\mu_{\ \nu}-b^\mu b_\nu,
\ee 
with $u_{\rm int}$ and $p=(\gamma_{\rm int}-1)u_{\rm int}$ representing the internal energy density and pressure of the 
gas in the comoving frame with adiabatic index, $\gamma_{\rm int}$, and $b^\mu$ - the magnetic field 4-vector
\citep{gammie03}. 
$R^\mu_\nu$ stands for the radiative stress-energy
tensor  and $n$ for the photon number density.

The magnetic field is evolved under the ideal MHD approximation,
\be
\label{eq.ind1}
F^{*\mu \nu}_{;\nu}=0,
\ee
where $F^{*\mu \nu}$ is the dual of the electromagnetic field tensor. The time
component of this equation reflects the no-monopole constraint, and the
space components correspond to the induction equation, which in coordinate basis takes the following form,
\be
\label{eq.Maxi}
\partial_t(\sqrt{-g}B^i)=-\partial_j\left(\sqrt{-g}(b^ju^i-b^iu^j)\right),
\ee
where $B^i$ is the magnetic field three-vector \citep{komissarov-99} which satisfies,
\be
\label{eq.Bit}
b^t=B^i u^\mu g_{i\mu}\ ,
\ee
\be
\label{eq.Bi1}
b^i=\frac{B^i+b^tu^i}{u^t}\ .
\ee 
The  
flux-interpolated contrained transport (Flux-CT) method of \cite{toth-00}
prevents numerical generation of spurious magnetic monopoles.

The radiative stress-energy tensor is obtained from the evolved
radiative primitives, i.e., the radiative rest-frame energy density,
and its four-velocity, $u^\mu_{\rm R}$, following the M1
closure scheme
 modified by the addition of radiative viscosity \citep[see][for details]{sadowski+koral,sadowski+dynamo}. 
The interaction between the gas and the radiation, i.e., the transfer
of energy and momentum, is described by 
the radiation four-force density $G^\mu$.
The opposite signs of this quantity on the right hand sides
of Eqs.~\ref{eq.cons2} and \ref{eq.cons3} reflect the fact
that the gas-radiation interaction is conservative, i.e., it transfers energy
and momentum between gas and radiation. The detailed form of the
four-force density is given and discussed in \cite{sadowski+electrons}.

\subsection{A new component in the electromagnetic stress-energy tensor}

Radiation generates an electric field if there is a non-zero radiation flux $F^{\hat i}\equiv G^{\hat i}/(\kappa \rho)$ in the fluid (comoving) frame (we denote with hats quantities measured in the comoving frame). Here, $\kappa=\sigma_{\rm T}/m_{\rm p}$.  This is, however, a necessary but not sufficient condition. We further need the plasma to consist of `light' electrons and `heavy' protons. In other words, the electric field results from the momentum transfer of the radiation pressure felt by the plasma electrons onto the plasma protons. Obviously, an electron-positron plasma may feel a non-zero radiation flux in its comoving gas frame, but no electric field develops. 

Within the formalism of \cite{sadowski+electrons}, the rate of momentum exchange between the gas and radiation is described by the spatial components of the radiation four-force, $G^\mu$.
As we have argued above, in the fluid (comoving) frame, for which $u^{\hat \mu}=(1,0,0,0)$, we obtain
\be
\label{eq.Ebat1}
E^{\hat \mu}_{\rm CB}=\left(0,\frac{\sigma_{\rm T} F^{\hat i}}{ec}\right)=\left(0,G^{\hat i}\frac{m_{\rm p}}{\rho ec}\right)\ .
\ee
We can rewrite eq.~(\ref{eq.Ebat1}) in covariant form (thus we can drop the hats) as follows,
\be
\label{eq.Ebat2}
E^{\mu}_{\rm CB}=h^{\mu}_{\nu}G^{\nu} \frac{m_{\rm p}}{\rho e c}\ ,
\ee
where $h^{\mu}_{\nu}=\delta^{\mu}_{\nu}+u^{\mu}u_{\nu}$ is the projection tensor, and $u^\mu$ is the gas four-velocity. The new components of the dual electromagnetic field tensor, therefore, are
\be
 F^{*\mu \nu}_{\rm CB}=-\epsilon^{\mu\nu\alpha\beta} (E_{\rm CB})_{\alpha}u_{\beta}=-\epsilon^{\mu\nu\alpha\beta} h_{\alpha\gamma}G^{\gamma}u_{\beta} \frac{m_{\rm p}}{\rho e c}\ ,
\ee
which leads to the modified induction equation
\[
\partial_t(\sqrt{-g}B^i)=-\partial_j\left(\sqrt{-g}(b^ju^i-b^iu^j) +
  \sqrt{-g}\epsilon^{ij\alpha\beta} (E_{\rm CB})_{\alpha}u_{\beta}\right)
\]
\be
\label{eq.induction2}
=-\partial_j\left(\sqrt{-g}(b^ju^i-b^iu^j+\epsilon^{ij\alpha\beta} h_{\alpha \gamma}G^{\gamma}u_{\beta} \frac{m_{\rm p}}{\rho e c})\right)\ .
\ee
It is interesting that the CB term involves the radiation four-force density $G^{\mu}$ which is already calculated in the numerical code, and does not require the calculation of extra quantitities. Therefore, its inclusion has no extra computational cost for an MHD+radiation simulation.

This is the first time that the extra CB term appears in the general relativistic form of the induction equation. Our goal in the present work is to include this term in an MHD+radiation numerical simulation and to thereby study its role in generating large scale magnetic fields in astrophysical black holes and accretion disks. It is important, though, to emphasize once again that the CB is a slow process, and this may have been one of the reasons why previous researchers did not include it in the induction equation. In fact, its inclusion would not modify the results of previous simulations since, in order to manifest its effect within the finite available computational time of those simulations, it would need to have been artificially amplified by several orders of magnitude.

\begin{figure}
	\begin{center}
\includegraphics[width=0.9\columnwidth]{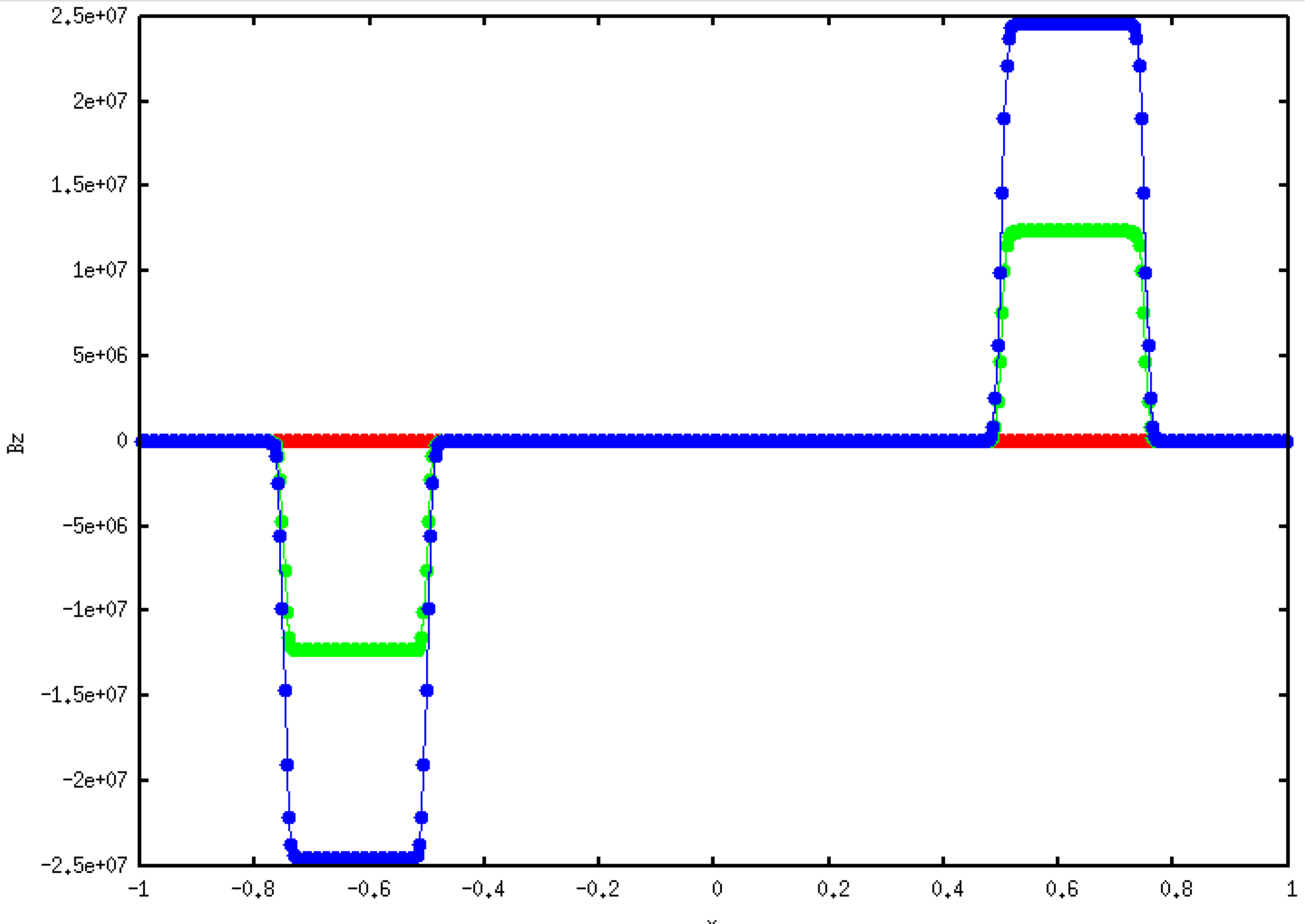}
	\end{center}
	\caption{$B^z~$ component of the magnetic field at the beginning of the test problem and at two consecutive times.}
  \label{CB0}	
\end{figure}

\subsection{Simple numerical test}
\label{s.verification}

Before we proceeded with a realistic simulation, we needed to check our numerical setup. We thus devised the following simple test. We considered a one-dimensional flow in the region $-L < x < L$,  and $-\infty < y < +\infty$, $-\infty < z < +\infty$. At time zero   ($t=0$) the fluid velocity is, $v = v^y = + V$. We further considered a simple radiation field with radiation flux along the $y$ direction in the lab frame such that\\
\vspace{1pt}\\
$f^y = 0$ for $x < -3 L/4$ and for $x > 3 L/4$,\\
$f^y = - F$ for $-L/2 < x < L/2$,\\
$f^y = - F  ( x + 3 L/4 )/ (L/4)$ for   $-3 L/4 < x < -L/2$,\\
$f^y = - F  ( x - 3 L/4 )/ (L/4)$ for $L/2 < x <  3 L/4$.\\
\vspace{1pt}\\
\noindent
We also considered an incompressible fluid with $\rho=\rho_0=\rm const$, and zero pressure ($p = 0$). Due to the radiation pressure, the velocity decelerates continuously as
$$v = v^y  = V +  ( f^y  \sigma_{\rm T} /m_{\rm p} )  t\ .$$
The radiation flux in the fluid frame becomes
$$f^{\hat y} = f^y  \Gamma^2 = \frac{f^y}{1-v^2/c^2 }\ .$$
This in turn generates an electric field in the fluid frame 
$$E^{\hat y}  = f^{\hat y}\sigma_T / e\ ,$$
the curl of which yields the growth of a magnetic field. Initially, the magnetic field and current are small, thus the resulting Lorentz force will not affect the velocity evolution derived above. 
As the fields grow though, they will affect the flow dynamics and the growth will saturate. At small times, $B = B^z = ({\rm d}f^{\hat y}/{\rm d}x)t/e$, thus\\
\vspace{1pt}\\
$B = 0$   for  $x < -3 L/4$ ,  $x > 3 L/4$ and  $-L/2 < x < L/2$,\\
$B = B^z \approx  4 Ft/Le$  for $-3 L/4 < x < -L/2$,\\
$B = B^z \approx -4 Ft/Le$  for $L/2 < x < 3L/4$.\\
\vspace{1pt}\\
Obviously, the magnetic field topology consists of vertical lines that close at infinite $z$ distance.

Our numerical code closely reproduces the above analytical magnetic field profile (see figure~1). We consider this result as a simple test of our numerical procedure, and are now ready to tackle the realistic astrophysical problem of a radiating magnetized accretion flow around an astrophysical black hole.

\section{The Cosmic Battery in optically thin accretion flows}
\label{s.battery}
\begin{figure}
	\begin{center}
\includegraphics[width=1.0\columnwidth]{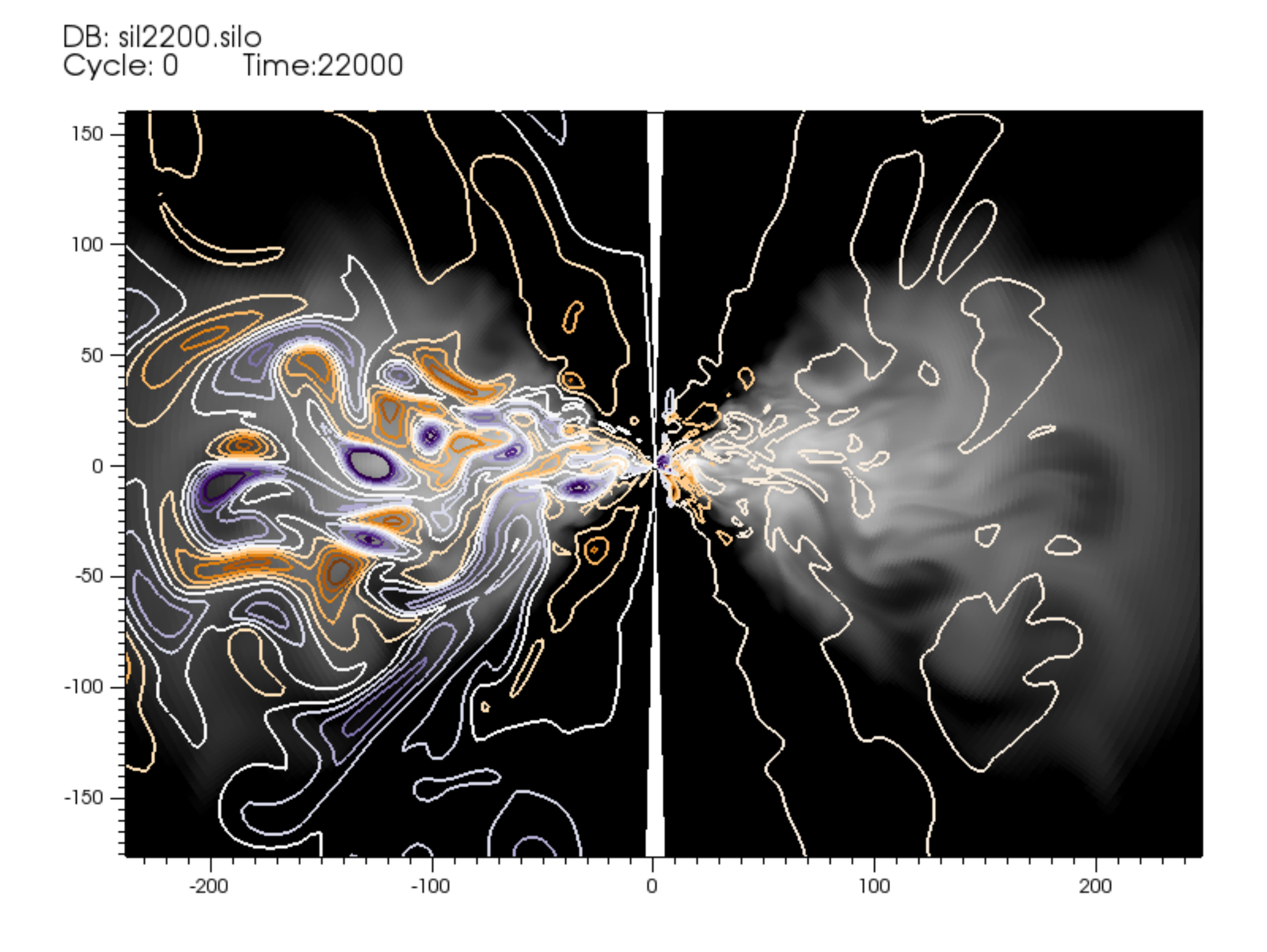}
\includegraphics[width=1.0\columnwidth]{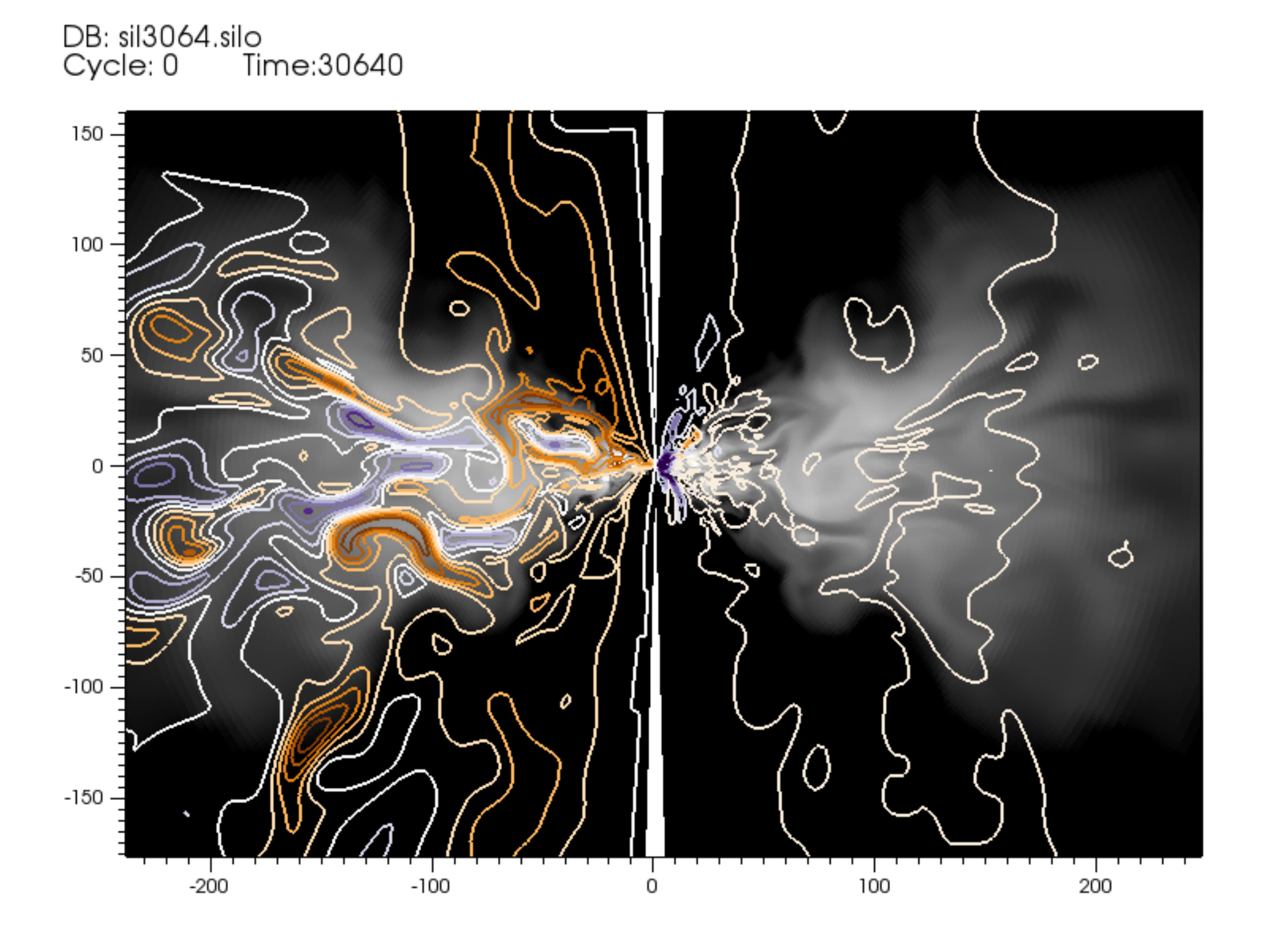}
\includegraphics[width=1.0\columnwidth]{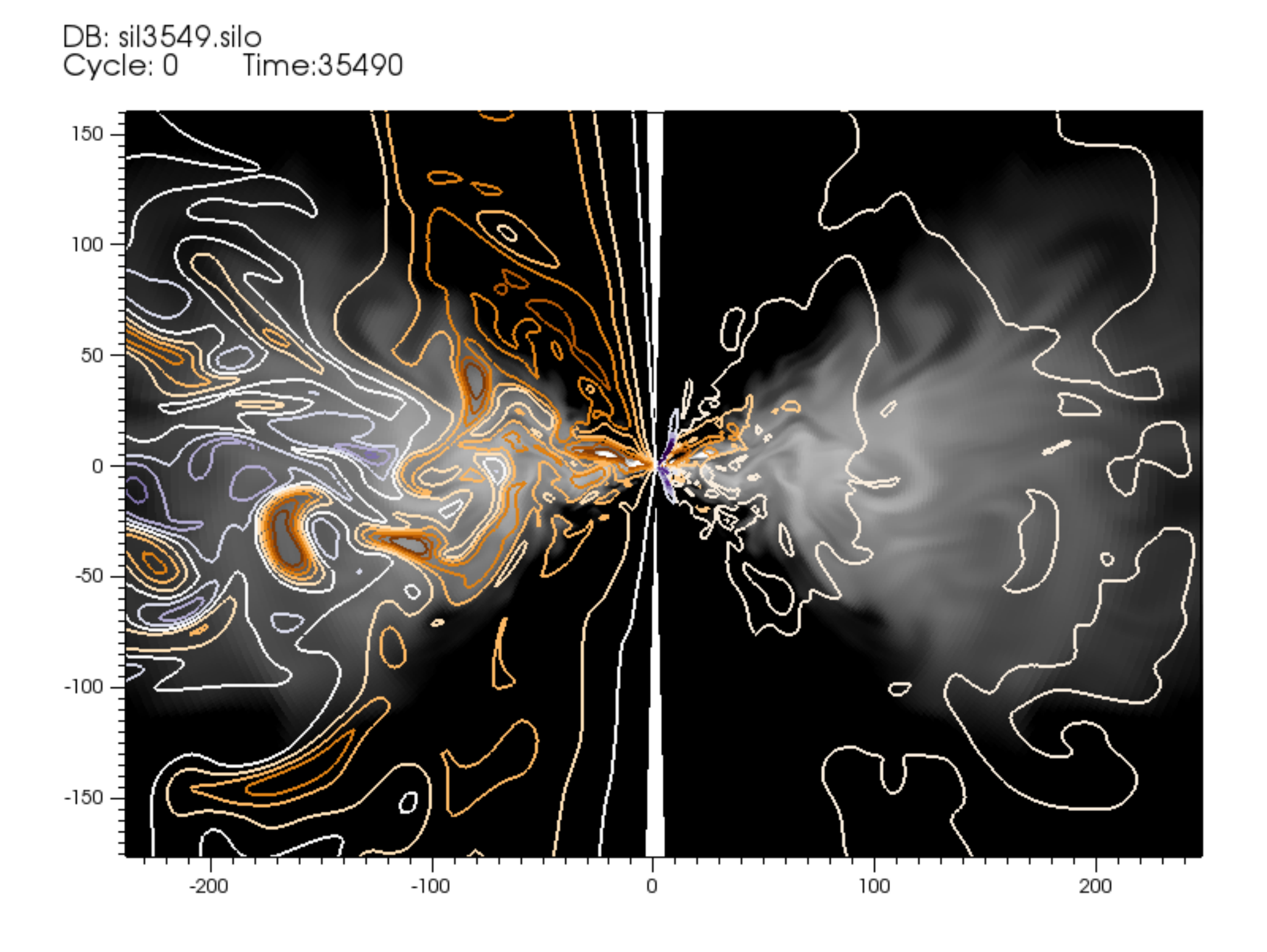}
	\end{center}
	\caption{Poloidal magnetic field evolution when the CB
	term is activated. In the left-hand panel we show the instantaneous poloidal field with orange and purple color denoting right and left-hand magnetic field loop direction with respect to the $\phi$ direction respectively. In the right-hand panel we plot the part of the induction equation due to the cosmic battery (last term in the r.h.s. of eq.~\ref{eq.induction2}). From top to bottom three 
	different time slices are shown, at $t= 22000, 30640, 35490 ~ \tau_g.$}
  \label{CB1}	
\end{figure}

\begin{figure}
	\begin{center}
\includegraphics[width=1.0\columnwidth]{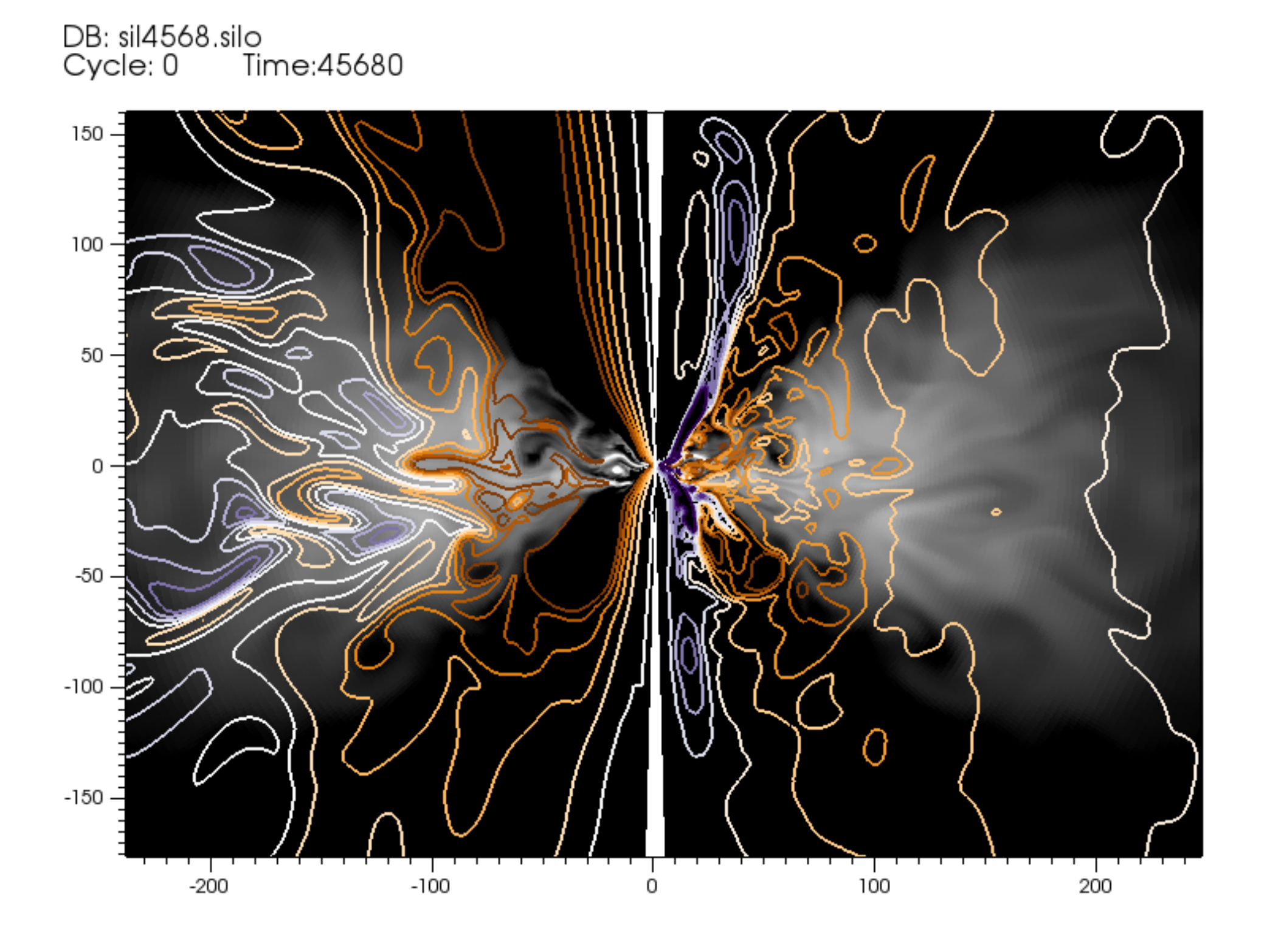}
\includegraphics[width=1.0\columnwidth]{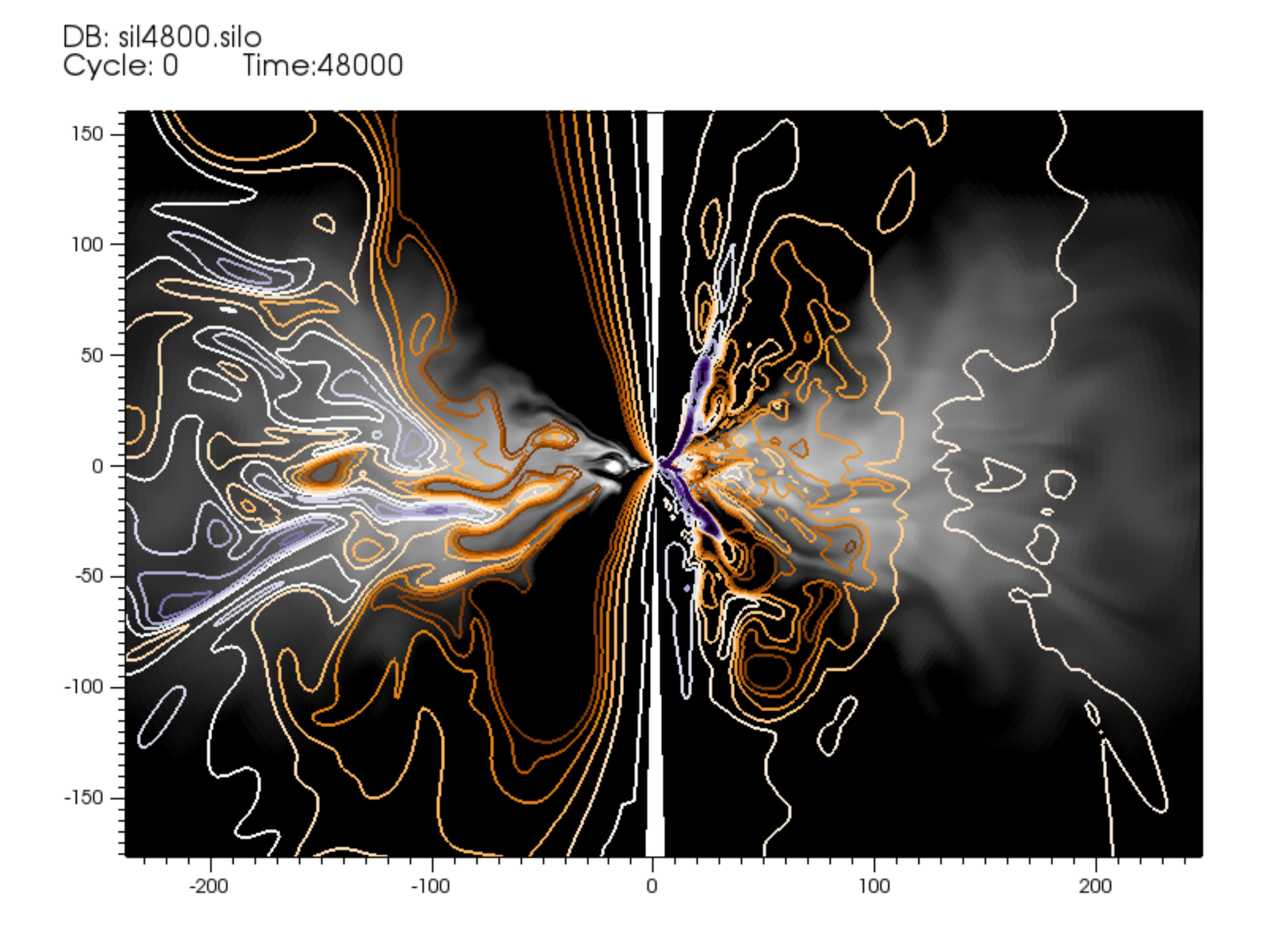}
	\end{center}
	\caption{Similar to figure \ref{CB1}, from top and bottom two
	different time slices are shown, at $t= 45680, 48000 ~ \tau_g.$}
	\label{CB2}
\end{figure}

We performed a set of two simulations of sub-Eddington, optically thin accretion flows around a non-rotating $10M_\odot$ black hole. The difference between the two simulations is the activation, or not, of the CB term due to the radiation field acting on the plasma electrons that we discussed in the previous sections. The aim of this comparison is to emphasize the importance of the CB term, since, as we will see,
it results in significant accumulation of non-zero magnetic flux
around the central black hole. Our results prove that the Cosmic Battery, if acting for a long enough time, may drive every accretion flow towards the magnetically arrested (MAD) state. Our simulations were performed in axisymmetry to reduce computational
costs. Snapshots of the magnetic field structure in the simulation
with the CB term at various times are shown in figures~(\ref{CB1}) \&
(\ref{CB2}). The left-hand panels show the
instantaneous poloidal field, whereas the right-hand panels show the
instantaneous contribution of the Cosmic Battery (last term in the r.h.s. of eq.~\ref{eq.induction2}). Orange and
purple color contours denote right and left hand side magnetic field
loops with respect to the $\phi$ direction respectively (equivalently,
these loops appear as counter-clockwise and clockwise in the left
panels of figures~(\ref{CB1}) \& (\ref{CB2}) respectivly, but
clockwise and counter-clockwise in the right panels respectively).

The initial setup is the same in both simulations. The torus is seeded with a purely poloidal magnetic field broken down into small poloidal loops of alternating polarity \citep[see][for details]{NSPK12}. The reason we consider a nonzero initial magnetic field is simply to trigger the MRI which will, in its turn, initiate accretion. Each loop carries the same amount of initial seed magnetic flux so that, without the CB term, the black hole is unable to acquire a large net flux over the course of the simulation. In other words, without the CB term, the accreting gas cannot and in fact does not become magnetically arrested, despite the very long duration of the simulation. This has been seen in detail in the `SANE' simulation of \citet{NSPK12} (`SANE' stands for `standard and normal evolution'), so there is no need to discuss the first part of the simulation where the magnetic field generated by the CB is still insignificant.

After $20000\ \tau_{\rm g}$ from the beginning of the `SANE' simulation, a second simulation was started in which we turned on the new CB term. We run the two simulations in parallel for another $25000\ \tau_{\rm g}$ and compared the two evolutions. As in the first simulation, the MRI continuous to generate turbulence, the disk continues to accrete, and the simulation proceeds for a while as a `SANE' simulation. Meanwhile, the CB generates poloidal magnetic field as seen in the right-hand panels of figures~(\ref{CB1}) and (\ref{CB2}), but other than that, there is no significant manifestation of the Cosmic Battery in the process. Nevertheless, as pointed out by \cite{CK98} and \cite{CCK08}, `the importance of the CB effect lies in its {\it secular nature}. The poloidal magnetic field is expected to be amplified by the continuous accumulation of magnetic flux due to the persistent large-scale velocity field, and its magnitude will eventually reach dynamically significant levels'. Indeed, at around $35000\ \tau_g$ the effect of the CB begins to manifest itself in the overall magnetic field structure!

As time progresses, the CB contribution becomes more and more significant. In a previous study, \cite{CKC06} and \cite{CCK08} simulated the CB-induced magnetic field growth in a Newtonian simulation under the assumption of a prescribed central isotropic radiation field. As a result, magnetic field loops were generated
mostly in the vicinity of the central object, namely at the position of the innermost stable circular orbit (hereafter ISCO). In the present study, though, the source of the radiation is the accretion flow itself, and therefore, the CB term is activated everywhere throughout the flow. Mostly poloidal magnetic fields are generated, and the poloidal component is subsequently converted to azimuthal by the shear of the flow. Magnetic field loops are generated not only close to the black hole in the ISCO neighborhood, but also further out, as seen in all right-hand side panels of figures \ref{CB1} \& \ref{CB2}. Obviously, the CB effect is stronger in the vicinity of the black hole.

In the middle panel of figure \ref{CB1} a large amount of magnetic loops with right hand side polarity with respect to the $\phi$ direction (orange color) are generated close to the black hole.
At this point the whole magnetic structure begins to diverge from a SANE picture with multiple magnetic loops of alternating polarity, to a structure dominated by big magnetic loops of one polarity in the region inside about $50 ~ R_{\rm g}$ around the black hole 
(orange color, middle panel fig. \ref{CB1}). Further out, the multiple polarity structure is still evident, but it is gradually wiped out as the simulation progresses. 

The first important result of the activation of the CB term is precisely the production of these magnetic loops. The inner foot point of such a loop ends, due to accretion, on the black hole. On the other hand, the outer foot point remains  anchored onto the rotating disk matter. This difference in rotation of the foot points of a magnetic field loop results in the loop puffing up and eventually becoming torn apart in two independent magnetic field lines. This opening up of poloidal magnetic field loops is a natural result seen also in simulations with a force-free 
environment above the disk and around the axis \citep[e.g.][]{CNK15, PGB15}. The final result of such a physical procedure, would be a large scale magnetic field of one poloidal polarity over the black hole, and a magnetic field of the opposite polarity threading the accretion disk. It is interesting that several specific predictions of the CB on the magnetic field polarities have been confirmed observationally \citep[see e.g.][]{CCKG09, CGKCKC16}. 

In the present ideal-MHD simulation magnetic field lines remain connected till the end of the run (bottom panel of figure \ref{CB2}). The outer part of each generated magnetic loop moves outwards through the disk, forming a fingered structure inside the disk as can be seen if the left-hand and bottom  panel of figure \ref{CB2}. In our present study, the outward movement of the outward footpoint is a result of numerical diffusion, since no explicit magnetic diffusivity 
has been included in the code. In realistic conditions, accretion disks are expected to be diffusive. We do expect that the introduction of magnetic diffusion in the outer parts of the accretion flow will have a significant impact on the growth rate of the magnetic field accumulated over the black hole \citep{CNK15}. Moreover, the magnetic field threading the disk at large scales will not reach equipartition values since it will continuously diffuse outward, probably also launching a magnetic disk wind in the process \citep{BP82, CL94}. Again, this intrinsic feature of the CB, namely the large scale outward field diffusion through the accretion flow, could be tested against observations as it comes for free to explain certain features of disk outflows \citep[e.g.][]{FKSBTC17}.

We measure the accretion rate  in our simulations
by integrating the mass flux over the black hole horizon.

\be 
\label{massacretionrate}
\dot{M} = \int_0^{\pi} 2\pi \sqrt{- g} ~ (\rho u^r) ~ d\theta\ , 
\ee
In both runs, accretion is due to the MRI, and the two accretion rates are of the same order. 
We further define $\Phi$ to be the normalized 
magnetic flux threading each hemisphere of the BH horizon and 
measure it by integrating the radial magnetic field over the BH horizon,
\be 
\label{magneticflux}
\Phi =\frac{1}{2 \sqrt{\dot{M}} } \int_0^{\pi}  2\pi\sqrt{- g} ~ |B^r| ~ d\theta\ , 
\ee

\begin{figure}
	\begin{center}
\includegraphics[width=1.0\columnwidth]{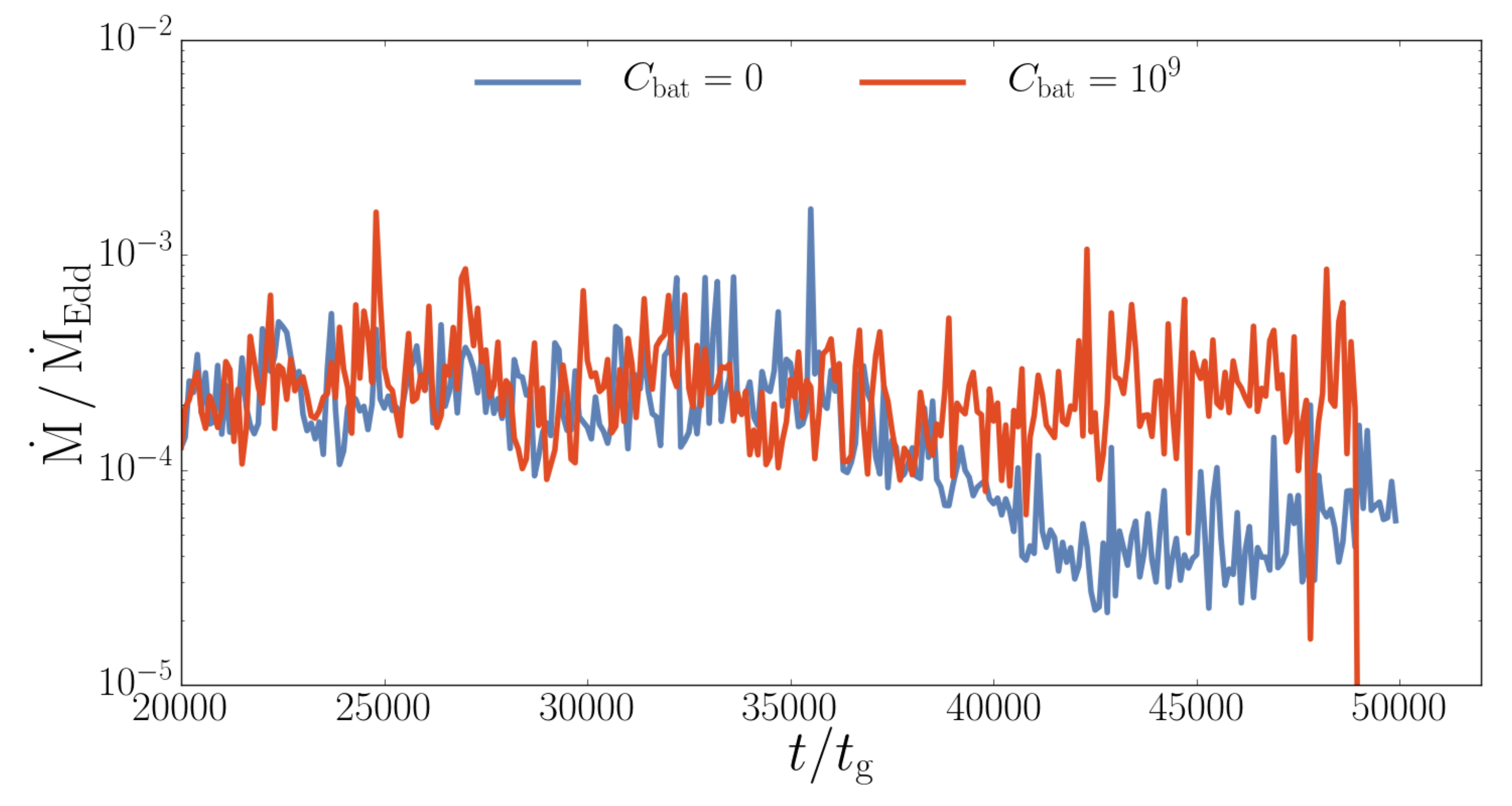}
\includegraphics[width=1.0\columnwidth]{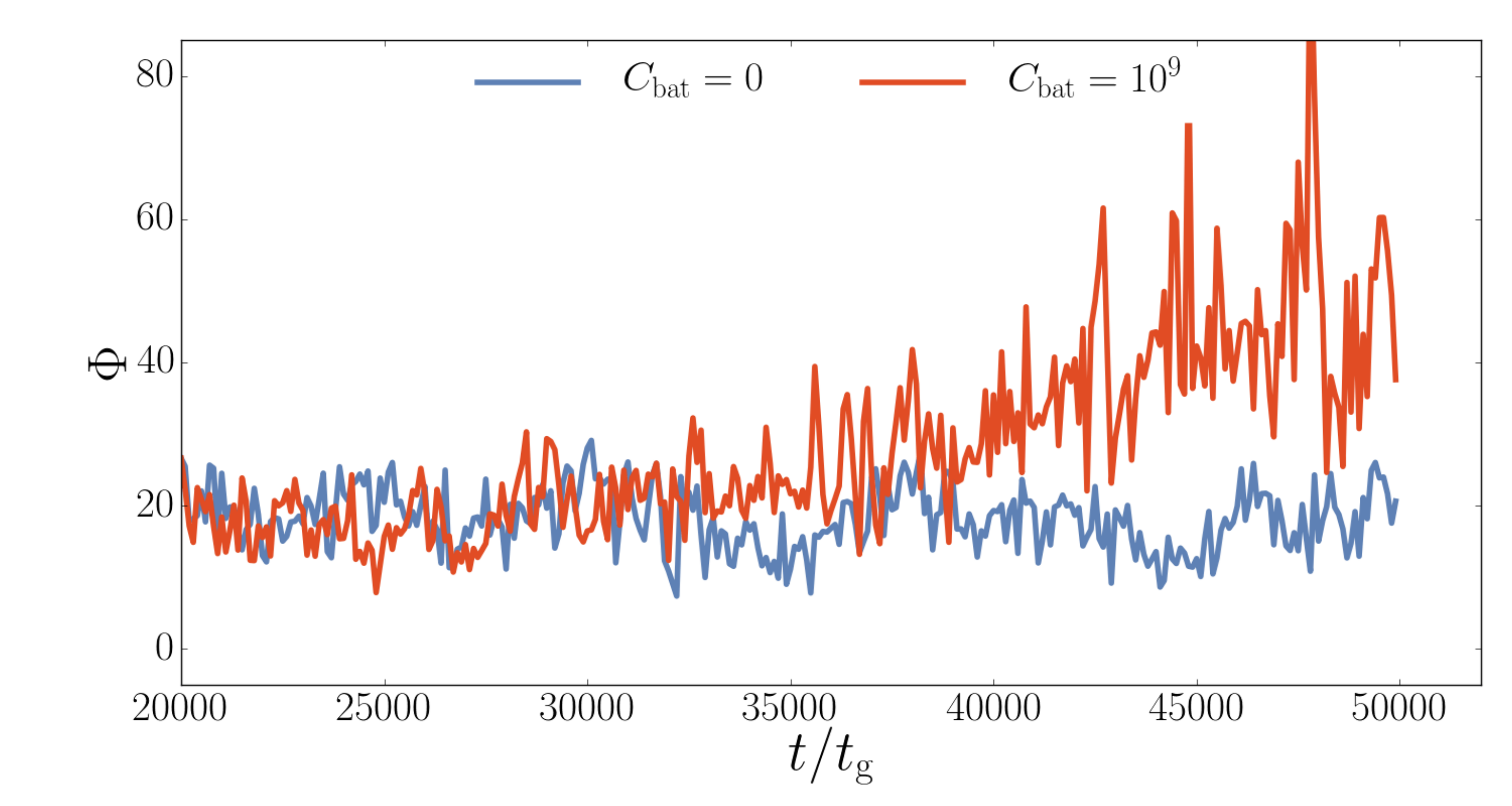}
	\end{center}
	\caption{Top panel: evolution of the mass accretion rate in the two simulations, one with the CB term amplified by a factor of $10^9$ (red line), and one without the CB term (blue line). Bottom panel: similarly for the magnetic flux parameter $\Phi$.}
	\label{flux}
\end{figure}

Interestingly, the CB mechanism does not saturate even in this ideal-MHD run, and continues instead to operates till the system becomes a magnetically arrested disk (MAD, \citealt{n03,b76,i03}). Notice that  the MAD state is obtained when the normalized magnetic flux reaches values $\Phi \sim 40-50~$ \citep{MAD10, MAD12}. The bottom panel of figure \ref{flux} shows a comparison of the magnetic flux between the CB run, where the battery term is activated, and a typical SANE evolution \citep{NSPK12}. After $35000\ \tau_{\rm g}~$ the evolution of the magnetic flux accumulation onto the black hole deviates from the typical SANE run and an almost linear growth is evident. 
We should state here once again that, in order to be able to observe this expected linear growth of the magnetic flux over a reasonable computational time, the CB term in our present run is amplified by a factor of $10^9$ .

\section{Summary and Discussion}

In this study we performed a comparisonn between two sub-Eddington, optically thin accretion flows, which were initially seeded with small scale poloidal magnetic flux of alternating polarity (SANE). In our basic CB run, the battery term discussed in sections \ref{s.introduction} and \ref{s.battery} was multiplied by a factor of $10^9$ to reduce computational cost, and was activated after $20000\ \tau_{\rm g}$ in a SANE simulation. The main result of the present study is that the initial SANE model is driven towards a MAD state. This is a strong indication that all 
accretion flows (from AGN to XRBs) may be driven by the CB towards a MAD state. This may take place once or several times during their lifetime, depending on the variable properties of the accretion flow which would directly affect their bursting activity \citep{KCKC12}.

The CB battery term gives rise to poloidal magnetic loops in the disk. As accretion proceeds, the inner part of these loops is advected onto the black hole horizon and contributes to the buildup of a large scale magnetic field of a particular polarity. We should state once again here the importance of the inclusion of physical magnetic diffusivity in the accretion flow. Obviously, allowing the outer part of the CB-generated magnetic loops to diffuse outwards has an important impact on the structure of the magnetic field that threads the accretion disk. 

We should also state here that the initial magnetic field configuration has no impact on the CB-generated magnetic field. A seed field was included just to initiate the MRI which subsequently leads to accretion. As we said, the CB is a secular process that operates continuously in the background of an otherwise turbulent flow. The CB can build a magnetic field from zero, i.e. even if the initial flow is unmagnetized. 

Our present numerical simulations were performed for a non-rotating (Schwarzschild) black hole. In that case, the radiation field inside the accreting flow is more or less isotropic, and therefore, in the frame rotating with the flow (the comoving frame), the radiation field is abberated in the direction opposite to the direction of rotation. This yields a radiation pressure force opposite to the direction of rotation, and this essentially corresponds to the well known Poynting-Robertson effect of the solar radiation acting on dust grains in the solar system. However, our previous work has shown that this may change as the spin of the central black hole is increased to more than about $70\%$ of maximal rotation. \citet{KC14} calculated the radiation field acting on the surface of a centrifugally supported disk-like flow distribution by performing general relativistic ray tracing from every other point on the surface of the disk. What they found was indeed unexpected. At low black hole spins, the abberated radiation field is qualitatively similar to what is described in eq.~(\ref{isotropicrad}). As the black hole spin increases, though, the inner edge of the disk (taken to lie at the position of the ISCO) approaches more and more the black hole horizon. As a result, the black hole horizon covers a larger and larger field of view of the flow at the inner edge of the disk, and as a result, the radiation field becomes strongly anisotropic. In addition, spacetime rotation induced by the black hole spin favors photon orbits along the direction of the black hole rotation, and not against it. \citet{KC14} discovered that at a black hole spin around $70\%$ of maximal rotation, the abberation of the radiation field vanishes. For higher black hole spins, the comoving azimuthal radiation force (thus also the CB induced electric field) changes sign near the black hole horizon since the innermost accretion flow is essentially illuminated only from behind\footnote{Obviously, the latter results apply only to prograde disk rotation (that is, rotation with the same direction as that of the black hole rotation). For retrograde rotation, the abberated radiation force is always opposite to the direction of rotation.}!

\citet{KC17} performed more detailed simulations of optically thin torus-like flow distributions (Polish doughnuts) and confirmed the conclusions of their earlier work obtained with a much simpler source of radiation (multi-temperature black body from the surface of the disk). Our next goal will be to reproduce these results for various black hole spins in a general relativistic MHD accretion flow simulation using the \koral\ code. If this change of direction of the CB effect around the inner edge of the accretion disk is indeed established, this will have strong implications for observations of large scale magnetic fields in astrophysical jets and disk \citep{CCKG09, CGKCKC16}. Finally, \citet{spin3} argued that the accumulation of a strong magnetic field around the central black hole may modify the position of the ISCO, and therefore, the change of direction of the CB effect around the inner edge of the accretion disk may also have implications on our observational estimates of black hole spin in active galactic nuclei and X-ray binaries \citep[e.g.][]{spin1}.

One more interesting question to explore in the future is whether the Cosmic Battery works well even when the accreting gas is optically thick. \koral\ can handle optically thick flows, so we believe we are capable of answering this question.
 
\section*{Acknowledgements}

AS acknowledges support for this work by NASA through Einstein Postdoctotral Fellowship number PF4-150126
awarded by the Chandra X-ray Center, which is operated by the Smithsonian Astrophysical Observatory for NASA under contract NAS8-03060. AN is supported by an Alexander von Humboldt Fellowship. RN was supported in part by NSF grant AST1312651 and by the Black Hole Initiative at Harvard University, which is supported by a grant from the John Templeton Foundation. The authors acknowledge computational support from NSF via XSEDE resources under grant TG-AST080026N.
\bibliographystyle{mn2e}

\end{document}